# Progress of the design of the 12-m optical, Wide-field Spectroscopic Telescope (WST)

P. Dierickx[a], G. Gausachs[b], C. Cudennec[a], D. Lee[c], R. Bacon[a], T. Travouillon[b], O. Bellido-Tirado[d], B. Sassolas[e], L. Pinard[e], S. D'Auria[f], C. Del Vecchio[f], V. De Caprio[f], V. Cianniello[f], W. Sutherland[g], P. Doel[h], T. Augusteijn[i], R. Navarro[i], L. Fréour[j], W. Saunders[k]
[a]Centre de Recherche Astrophysique de Lyon, France; [b]Australian National University, Australia; [c]Science and Technology Facilities Council (STFC), United Kingdom; [d]Leibniz-Institut für Astrophysik Potsdam, Germany; [e]Laboratoire Matériaux Avancés- IP2I, CNRS UCB Lyon, France; [f]Istituto Nazionale di Astrofisica, Italy; [g]Queen Mary University London, United Kingdom; [h]University College London, United Kingdom; [i]NOVA optical infrared instrumentation group at ASTRON, Netherlands; [j]Universität Wien, Austria; [k]Astralis - MQ

## ABSTRACT

The Wide-Field Spectroscopic Telescope (WST) is a 12-m class, Alt-Az optical telescope providing two simultaneous foci, 2° and 3x3 arc minutes respectively, for Multi-Object Spectroscopy (MOS) and Integral Field Spectroscopy (IFS) at visible wavelengths. Within the framework of the Horizon Programme, the project has been approved for phase A funding by the EC in 2024 and is currently undergoing intensive development. The optical design relies on a corrected Cassegrain solution feeding multi-object spectrographs through fibres, while the central area of the field is propagated to a gravity-stable Integral Field Station housing an array of spectrographs. We report progress of the design, of the functional and physical definitions of key subsystems, and of the design of high-performance coatings of the telescope optics. The telescope would be located near ESO's Paranal-Armazones observatory and protected by a VLT-like enclosure. With a view to minimising risks, from capital investment to operations, designs build extensively on VLT and ELT solutions. Owing to the optical configuration, the telescope can be wavefront-controlled at both foci, with little if any crosstalk between MOS and IFS controls. The optical path is constrained to provide natural guide star, ground-layer adaptive correction at the IFS field, within existing technology. Finally, from its earliest stage of development the system is being scrutinized for its properties in relation to sustainability.

## 1. INTRODUCTION

The Wide-field Spectroscopic Telescope (WST)[1],[2] is a 12-m class optical, wide-field telescope providing multi-object (MOS) and integral field (IFS) spectroscopy. The concept is currently under phase A development by an international consortium of institutes. The project started early 2025 after being awarded funding from the European Union within the framework of the Horizon Europe Research and Innovation Action. The reference design is under development, with key functional and physical characteristics mostly stable. The project is now approaching mid-term review, scheduled for Q3/2026. The telescope and enclosure level 1 requirements were released in July 2025. The reference optical design has been selected and analysed; a preliminary physical architecture is available; the functional role of each product is being outlined; the telescope structure and dome are being iterated and first finite element and CFD analysis have been performed. The design of the wavefront control strategy has just started, concurrently with engineering budgets (e.g. optical performance, presetting overheads, etc).

The objective of phase A is not to develop a fully optimised solution but a plausible, affordable point design, explore the design space, understand the risks and constraints, identify plausible improvement paths, and pave the way for design and construction. 10-m class, wavefront-controlled segmented telescopes already exist; none of them has an étendue comparable to that of WST, none of them operates two foci at the same time.

The telescope is required to deliver two *simultaneously available* focal surfaces (Figure 1): a Ø 2º focal surface for fibre-fed Multi-Object Spectroscopy (MOS), up to ~32,000 fibres, and a 3 x 3 arc minute$^2$ focal surface, located anywhere within a Ø 13 arc minutes patrol field, for Integral Field Spectroscopy (IFS) at a gravity-stable station. The MOS low- and high-resolution instrument, and the IFS instrument, are described elsewhere[3][4],[5][6]. Some level of obscuration ("dead zone") between the MOS and IFS sky areas is tolerated (Figure 1).

The above summarises the originality but also the complexity of the system:

1º Large collecting power and MOS field of view implies unprecedented étendue;
2º Fibre-fed spectroscopy implies a fast focal ratio (~f/3) at the optical interface between telescope and fibres;
3º Integral Field Spectroscopy with image slicers implies a long IFS focal ratio (~f/20 to f/50);
4º As the 3x3 arc minutes2 IFS field may be offset anywhere in a Ø 13 arc minutes one, lines of sight of the MOS and IFS fields are not necessarily aligned.
5º Although the size of the IFS field may seem modest for imaging, it shall be noted that it is three-dimensional; the complexity and volume of the instrument could – roughly – be characterised by the product R x N, where R is the spectral resolution and N the total number of spaxels (or field area). The implied volume of the instrument is considerable to say the least, and calls for fixed, gravity-stable environment;
6º Simultaneous operation of two fields in a 12-m class telescope implies that within the altitude range and envelope of environmental conditions, both focal surfaces must be cross-registered and actively held to specification in real time.

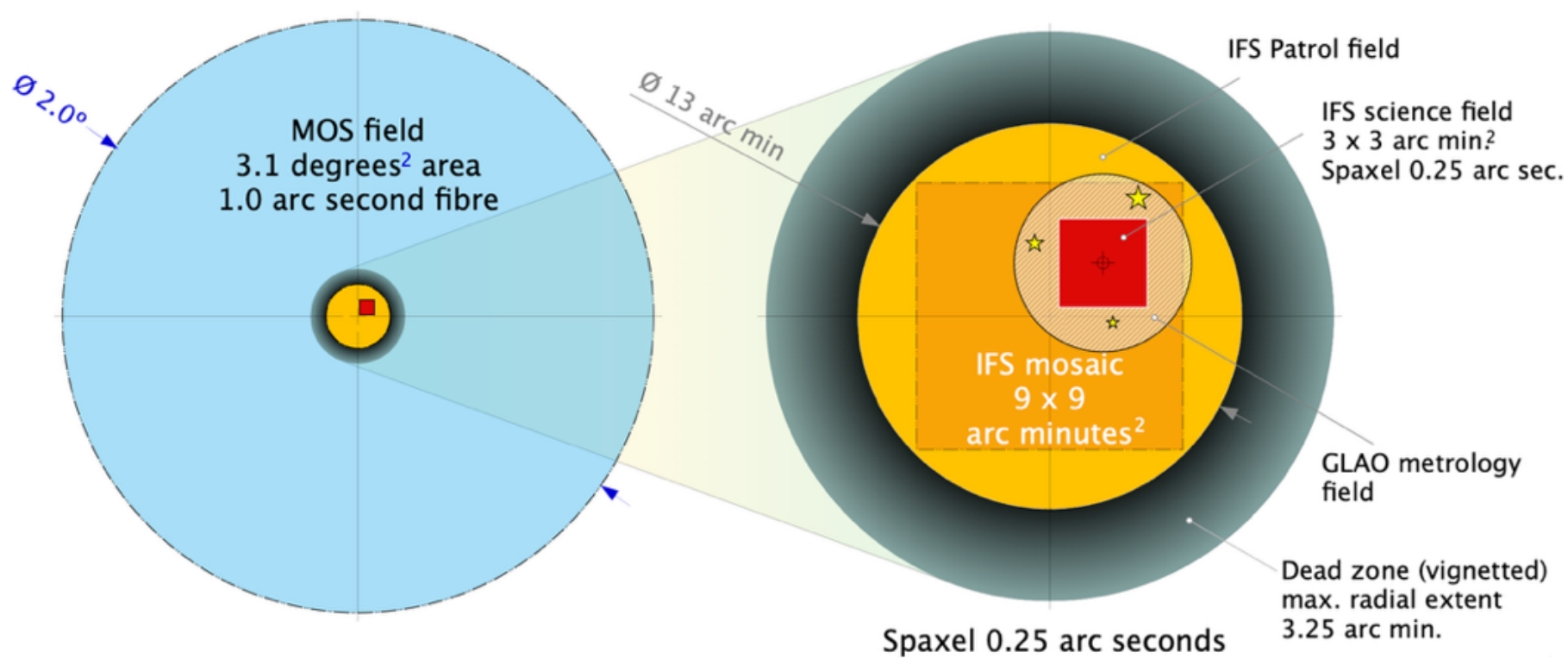

Figure 1 – MOS and IFS fields, notional representation

The telescope shall provide for seeing reduction (ground layer adaptive optics) at the IFS focus. The design therefore includes a conveniently conjugated reflecting surface of suitable dimensions and provides observables over a control field of view sufficiently large for reasonable sky coverage with natural guide stars. The field of view propagated down to the IFS station has therefore been set to Ø 6 arc minutes, without vignetting over the 3x3 science field and only minor vignetting allowed outside.

An essential guideline is to contain high-level departure from classical designs, despite the unusual dual-foci requirement. In a nutshell, the telescope itself should not look too different from the current generation of 8- to 10-m class telescopes, allow similar control principles, and be housed in a familiar-looking enclosure. Even though such guideline is prone to subjective interpretation, it has been vigorously applied to high level trade-offs.

At lower level, the design strategy for the telescope subsystems calls for low-risk, cost-efficient technological solutions. Wherever possible, opto-mechanical assemblies resemble existing solutions, including but not necessarily limited to ESO's Extremely Large Telescope (ELT)[7] ones. The primary mirror is made of 78 segment modules identical to ELT ones except for their optical prescription. Sensors, actuators, support systems are presumed to have same specification as the ELT ones – even though flexures will be considerably lower and thus stroke requirements less demanding. The adaptive mirror for seeing reduction at the IFS focus is similar to the adaptive secondary mirror of ESO's Very Large Telescope[8], in shape and in size. All refracting components are made of Silica and constrained to not exceed 1.6-m clear aperture. Reflecting ones are presumed to be made of glass or glass-ceramic (e.g. Zerodur™ or ULE™) and not exceed 4.0-m clear aperture; any fast-steering mirror shall have dimensions comparable to the ~2.5m size of the ELT M5 and preferably be made of glass or glass-ceramic.

The system being conceived as a survey telescope, optical efficiency (quality, throughput) and availability – including minimal overheads – are essential performance criteria. The high-level metrics for optical quality ω is the energy concentration in the fibre (MOS) or spaxel (IFS), normalised to the energy concentration, which would be obtained if the sole perturbation were the free atmospheric turbulence. The instantaneous optical efficiency E of the system is the product

$$E = A.\omega(\vec{r}, \Delta\lambda, \theta_0, z).\tau(\lambda).V(\vec{r}), \qquad (1)$$

where A is the collecting area, τ is the throughput, V is the vignetting, $\vec{r}$ represents the field coordinates, Δλ is the wavelength range, $\theta_0$ the seeing at airmass=1, and z the zenithal distance.

The target is to maximise the telescope efficiency E over the wavelength range 0.37-1.60 μm (MOS), respectively[1] 0.37-0.93 μm (IFS), from z=0 to 60º, within excellent seeing conditions ($\theta_0$=0.5 arc seconds at λ=0.5 μm, airmass=1). The target for the MOS optical quality is $\omega \geq 0.90$ over ~80% of the fibres, including field aberrations, system-level contingency, and all perturbations excluding free atmospheric turbulence. This corresponds to ~0.17 arc seconds RMS intrinsic quality of the telescope (i.e. nominal field aberrations excluded), down to MOS focal surface. At the IFS focus, a precise optical quality budget is still to be consolidated, but we already expect higher optical quality than at the MOS focus. The reduced field of view and the availability of additional (weakly) aspherical surfaces allows for excellent as-designed optical quality; we also anticipate that part of the stroke of the adaptive mirror will be allocated to mitigating telescope perturbations, including local air turbulence and wind buffeting on the primary mirror.

Throughput is being assessed[9], under the assumption that all surfaces not exceeding ~1.6-m could receive durable dielectric coatings optimised over the corresponding wavelength range (0.37-1.60 μm for the MOS optics, 0.37-0.98 μm for the IFS optics). This may include the primary mirror segments, insofar as

a) the temperature of the segments upon coating would not exceed limits deemed safe for the bonded interface pads,
b) the strain resulting from surface stresses would remain within the correction range of the warping harnesses, and
c) periodic cleaning, if possible in-situ, could be an acceptable alternative to recoating over a system lifetime of 30 years – admittedly, a challenging proposition.

This would also save on operational costs and capital investment – spare segments would no longer be required.

Assuming a combination of freshly deposited, dielectric coatings and Vera-Rubin ones[10] for the largest mirrors (i.e. exceeding 1.6-m), the theoretical throughput is estimated

- For the MOS, at 88% average and 77% minimum over the bandwidth 0.37-1.6 μm
- For the IFS, at 71% average and 61% minimum over the bandwidth 0.37-0.98 μm

These figures may eventually prove conservative. GRaded INdex (GRIN) coatings are of particular interest for the antireflection coatings of the refracting optics. Combining Vera Rubin-type coatings for the largest mirrors, dielectric ones for reflective surfaces smaller than 1.6-m, and GRIN anti-reflection coatings for the refracting ones, would theoretically result in

- For the MOS, 91% average and 83% minimum over the bandwidth 0.37-1.6 μm
- For the IFS, 80% average and 71% minimum over the bandwidth 0.37-0.98 μm

In the context of astronomical applications, however, our understanding is that the TRL (Technology Readiness Level) is low. A development programme is therefore being undertaken by the Laboratoire des Matériaux Avancés (LMA), part of the WST consortium[9].

A 12-m class telescope inevitably requires wavefront control to meet specification over the typical length of an observing block. The system design, including the optical solution, is constrained to allow simultaneous wavefront control without significant crosstalk between the MOS and IFS loops. The optical path to MOS focus is – in good approximation – common path; the telescope can therefore be "driven" by the sky metrology at the MOS focal surface, while the IFS-specific path includes the sky metrology and degrees of freedom necessary to cross-register to the MOS line of sight and manage differential perturbations. In view of the large field of view, and taking into account, first, that the fibres will not relocate during an observing block, and second, that the segmented primary mirror is subject to lateral expansion of its supporting steel structure, care must be taken that the control strategy complies with stringent requirements as to plate scale stability over the length of an observing block.

---

1 All IFS properties presented herein apply with a slightly more demanding requirement 0.37-0.98 μm, which has recently been relaxed to 0.37-0.93 μm.

Several sites, all located near Paranal Observatory, are under consideration. Pending final selection of the site, the WST design assumes same environmental conditions as applicable to Paranal, and same operational limitations.

Regarding sustainability and its role in design trade-offs, at the time of writing of this article the effort[11] focusses on long-term operation of the instruments and on science operations, as these were identified as key areas for mitigation during the design phase. We also expect the construction of the telescope and the cooling of the dome to contribute a fair amount to the total carbon footprint, due to the amount of concrete required for the telescope's base and the energy consumed to cool and stabilise the dome's temperature.

## 2. TELESCOPE OPTICS

The design of the MOS configuration is a corrected f/3.3 Cassegrain, with a 3-lenses corrector and the 2º focal surface above the primary mirror. Field curvature is approximately concentric with the exit pupil. The f/0.95 primary mirror is hyperbolic, the 2.5-m secondary mirror an 8$^{th}$ order asphere. The contour of the 78-segments primary mirror has minimum and maximum outer dimensions of 11.1 and 12.6-m, respectively, inner 2.8 and 3.8 m. The corrector includes aspherical lenses and provides atmospheric dispersion compensation at non-negligible zenithal distances.

A Ø 6 arc minutes field is extracted anywhere within the central 13 arc minutes of the MOS focal surface, by way of a pair of flat folding mirrors (M3 and M4), sent to a collimating mirror M5 on the altitude axis, then through a Schmidt plate towards a Nasmyth relay and eventually, through a coudé relay, down to a gravity-stable, telecentric, flat f/28.5 focal surface, coaxial with the altitude axis. The coudé relay also performs optical field de-rotation for the IFS field.

The optical layout is shown in Figure 2. The volume between the corrector and the MOS front-end is optically closed by the Schmidt plate and could therefore be sealed against dust contamination of the fibre positioners. An alternative, simpler optical solution has been explored, merging the Schmidt plate and the collimating M5 mirror into a single Mangin mirror; it delivers similar optical quality and reduces crowding in said volume, but would leave the fibre front-end exposed to dust contamination.

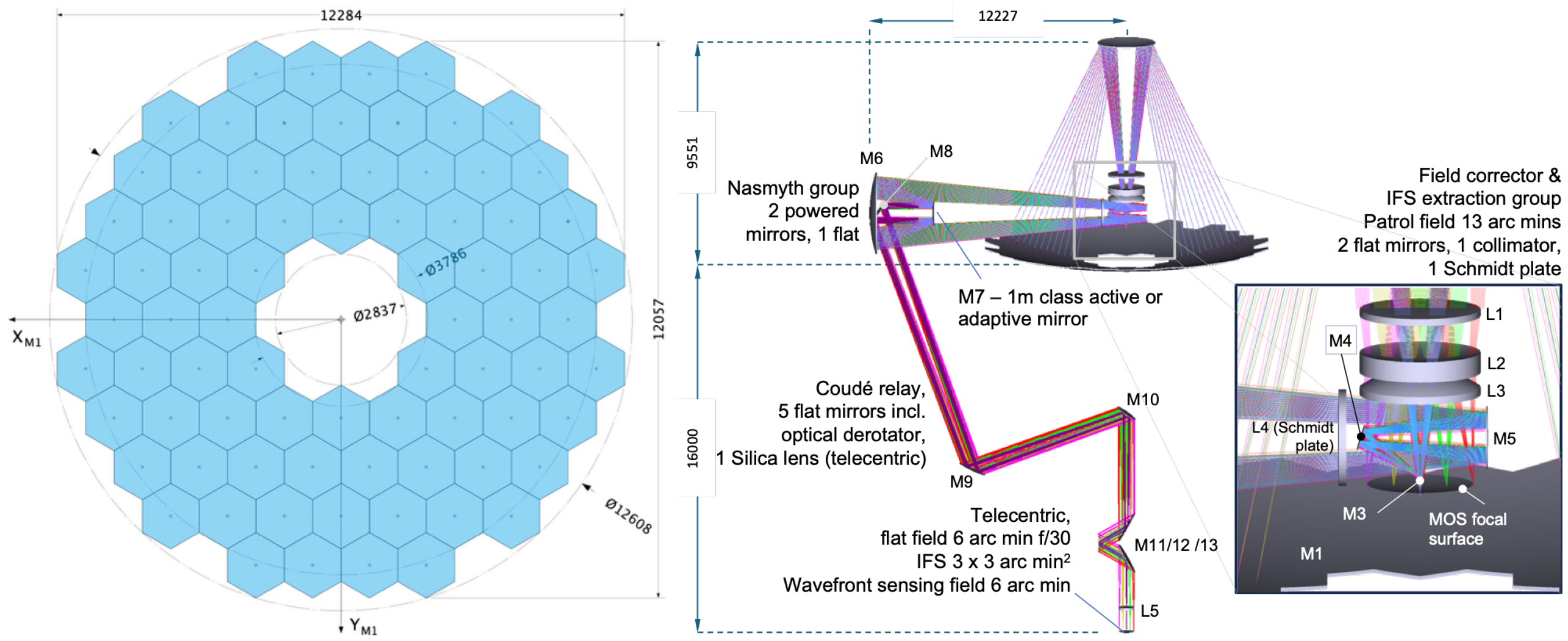

Figure 2 – WST primary mirror geometry, telescope optical layout

Extraction of a 6 arc minutes IFS field anywhere in the central 13 arc minutes area of the MOS focal surface is performed by the pair M3-M4. Both mirrors are flat, with elliptical contour (major axis 225, respectively 356 mm). The required degrees of freedom are piston and tip-tilt for mirror M3, and tip-tilt for M4. Mirror M3 being located above and close to the MOS focus, needs to be mounted onto the fibre front-end. This has the undesirable consequence of implying a (small) M3 rotation stage, to cancel the field de-rotation of the front-end and maintain mirror M3 pointing towards M4. The latter is mounted on the altitude of the telescope. The Nasmyth group of relay optics includes a Ø 3.6-m M6 mirror (f/0.85 conic surface, with moderate aspheric departure 410 µm), a convex, Ø 1.1-m aspherical mirror M7 (f/1.1, 84 µm aspheric departure), and a flat folding M8 (elliptical 1250 mm long axis), located near M6 and supported through its central hole. Mirror M7 is conjugated to ~100 m above the primary mirror i.e. ideally located for ground-layer adaptive optics.

Atmospheric dispersion compensation is provided by decentres of the last lens within the field corrector, combined with decentres of the secondary mirror. The optical solution has been optimised to mitigate variations of the asymmetrical distortion pattern with changing zenithal distance and atmospheric refraction. The worst-case displacement of the polychromatic spot centroids at the MOS is found to be about 0.38 arc seconds within 20 minutes observing block, starting from a zenithal distance of 50º. At the IFS, it is 0.06 arc seconds.

The as-designed optical quality at MOS focus as a function of the fibre population, is shown in Figure 3, at zenithal distances ZD=0, 30, 60º. The underlying design of fibre front-end has a hexagonal contour[12], with a tip-to-tip dimension of 2º on-sky. The additional allocation for all perturbations other than telescope nominal field aberrations and free atmospheric turbulence is 0.17 arc seconds RMS, including contingency – the latter being managed at system level.

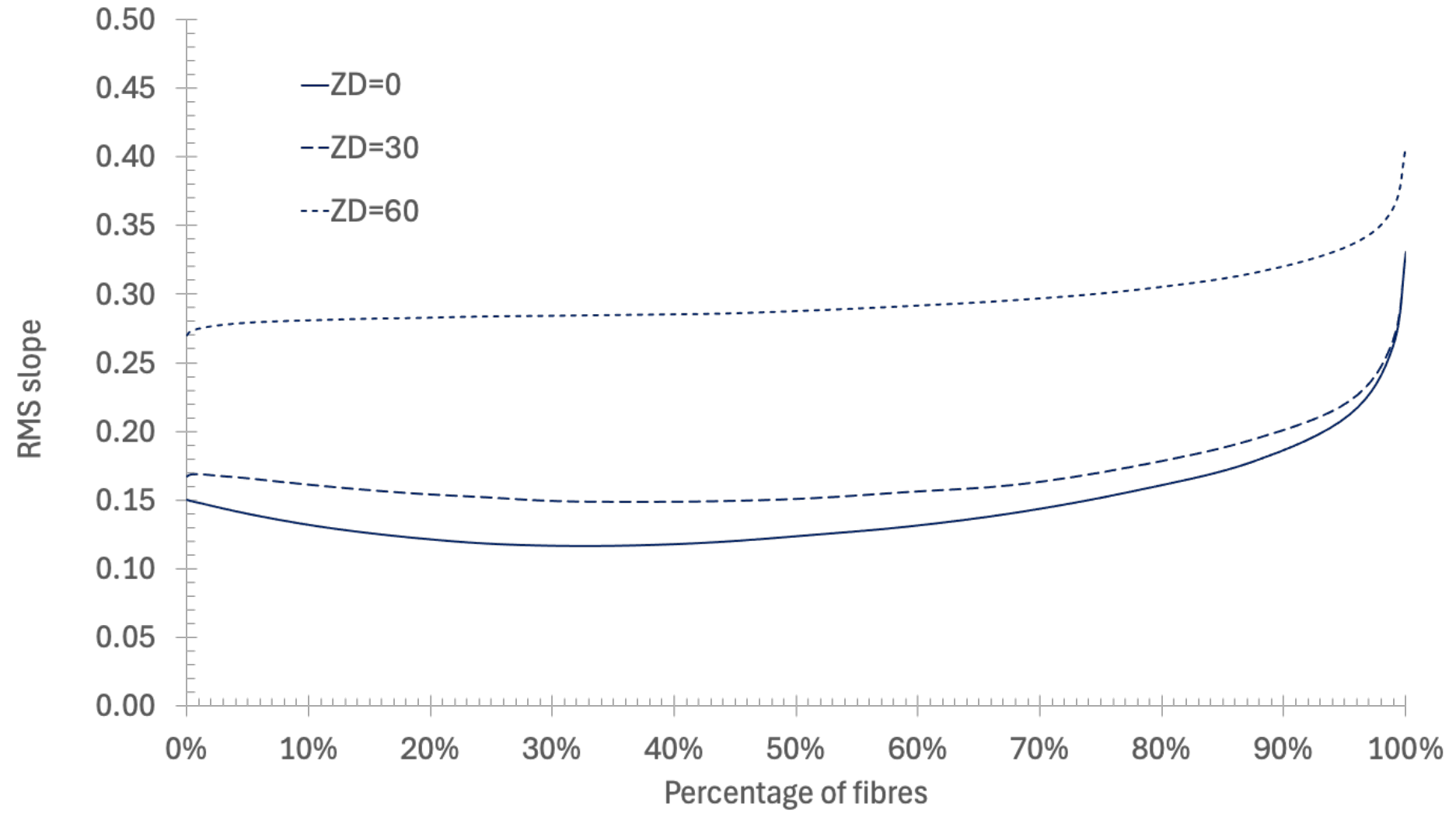

Figure 3 – Polychromatic RMS spot size, as-designed, MOS focal surface. ZD is the zenithal distance.

With a smaller field of view and powered surfaces providing additional degrees of freedom for optimization, the optical quality at the flat IFS focal surface is excellent, with RMS spot size ranging from 0.02 to 0.11 arc seconds, depending on zenithal distance and sub-field position within the 13 arc minutes IFS patrol field (Figure 4). The target is however more demanding than at the MOS focus: first, the spaxel size (0.25 arc seconds) is small compared to the fibre one (1 arc second); second, the reference Point Spread Function is the reduced seeing permitted by ground-layer adaptive optics; third, the telescope itself is only part of an extended error budget, which also includes all instrument perturbations.

Baffling is driven by the wide field and dimensioned to prevent undesired illumination of the fibres, at the inevitable cost of vignetting. The secondary mirror and the Cassegrain baffles clip the inner contour of the primary mirror, as seen on-axis from the MOS focus. The baffles need to be retractable for on-sky phasing calibration of the edge sensors of the primary mirror segment modules according to established pupil registration techniques[13], [14]. However, design space is extremely tight for the Cassegrain baffle; rock-solid reliability is mandatory; folding mechanisms (of e.g. baffle petals) are bound to result in low stiffness, and local modes could be excited under wind buffeting. However, different phasing techniques have been pioneered with ESO's APE experiment[15] and development within the framework of the ELT and Thirty Meter Telescope (TMT) project is continuing[16],[17]. We submit that the odds of this undesirable feature becoming unnecessary are reasonably high.

The total collecting area, as seen on-axis, is 97.3 $m^2$ at MOS focus, and 92.0 $m^2$ at the IFS one. MOS surfaces have been dimensioned to provide maximum visibility of the primary mirror on-axis; the effective area accepted by the fibres may be lower, depending on their numerical aperture. At the IFS focus, said visibility is no longer required and the surfaces have been dimensioned for 3 x 3 arc minutes unvignetted, with a 12-m circular pupil. The difference stems from the fact that the IFS instrumentation clips the pupil at 12-m anyway, and has a slightly higher central obscuration than the MOS one. The reason for the latter is understood; the minor update required to eliminate the difference has been identified.

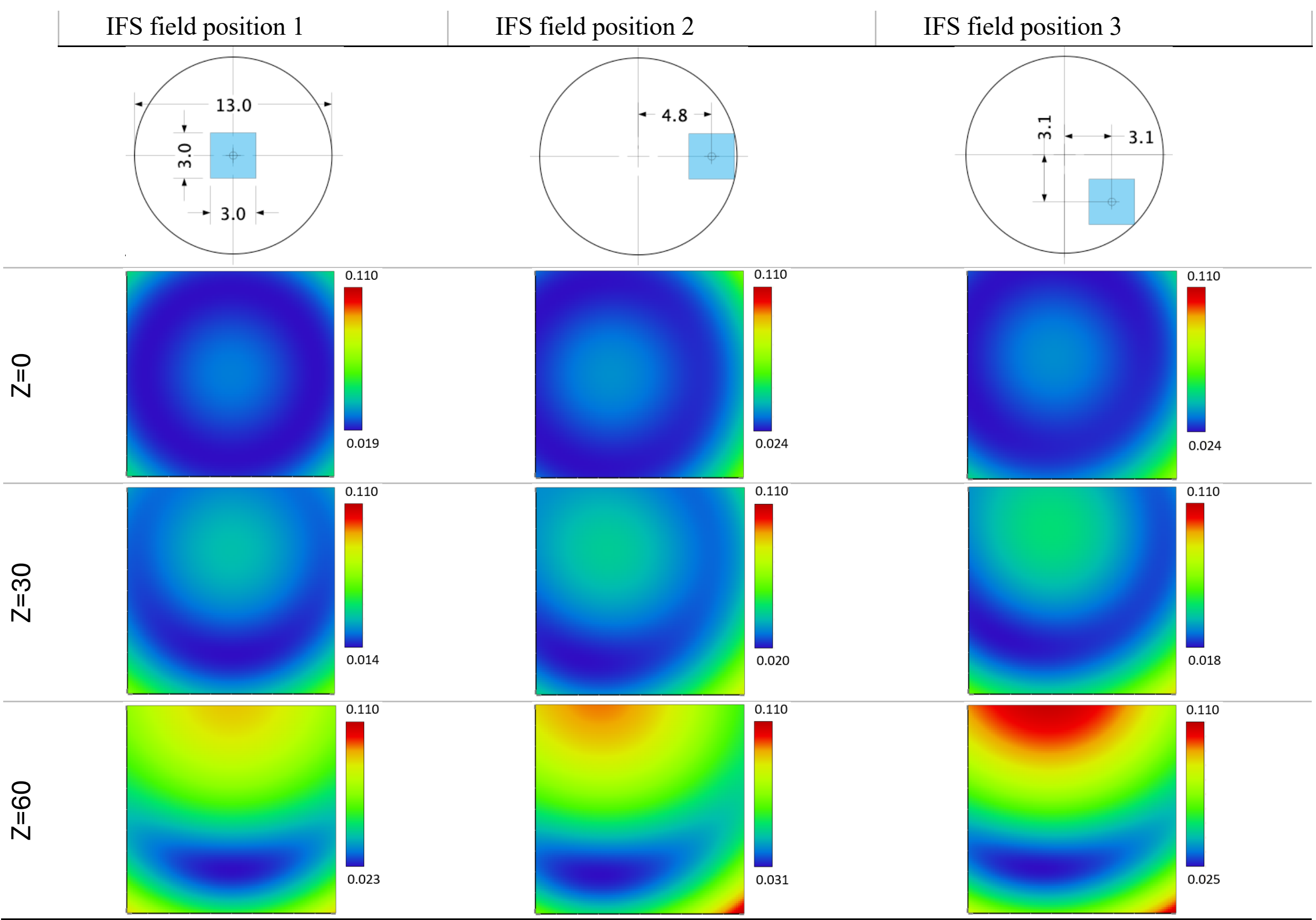

Figure 4 – IFS focal surface, maps of spot radius RMS, zenithal distance 0, 30, 60º. Units: arc seconds. Each box is 3x3 arc minutes[2]; the top row shows the location of the box within the patrol field.

Additional vignetting off-axis is small (up to ~5% loss compared to on-axis) up to ~0.9º off-axis, and set by the secondary mirror, whose clear aperture has been limited to 2.56-m. Beyond 0.9º, vignetting by the first corrector lens (Ø 1.6-m) kicks in, increasing losses by up to ~12% (compared to on-axis) at the very edge of the field of view.

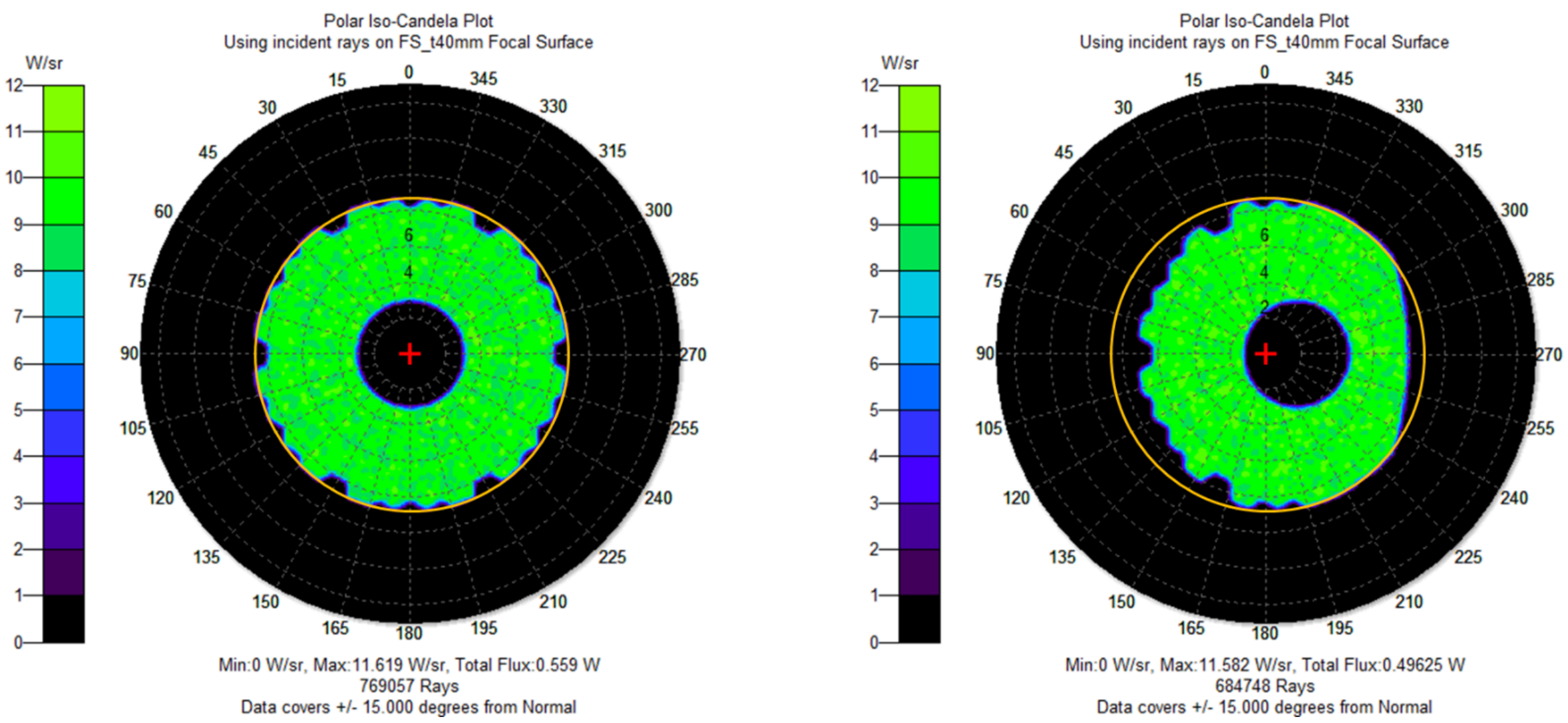

Figure 5 – Polar Isa-Candela Plot to show the angular distribution of rays arriving at the MOS focal surface for illumination with an on-axis point source (left) and 1º off-axis. The red cross indicates the centre of the polar plot. The orange circle indicates a cone half angle of 8.6º

Fibre illumination, considering the fibre acceptance angle, clear apertures, exit pupil position and field curvature, is illustrated in Figure 5.

The design has been scrutinized for stray light, ghosts, and pupil concentration (pupil ghosts). The most prominent ghost image identified occurs for an on-axis point. This on-axis pupil ghost can be seen in Figure 6 (MOS focus). The intensity of the pupil ghost is $1.2 \times 10^{-6}$ relative to the intensity of the on-axis point source. This is considered negligible and conservative as the model assumes un-coated refracting surfaces in the field corrector.

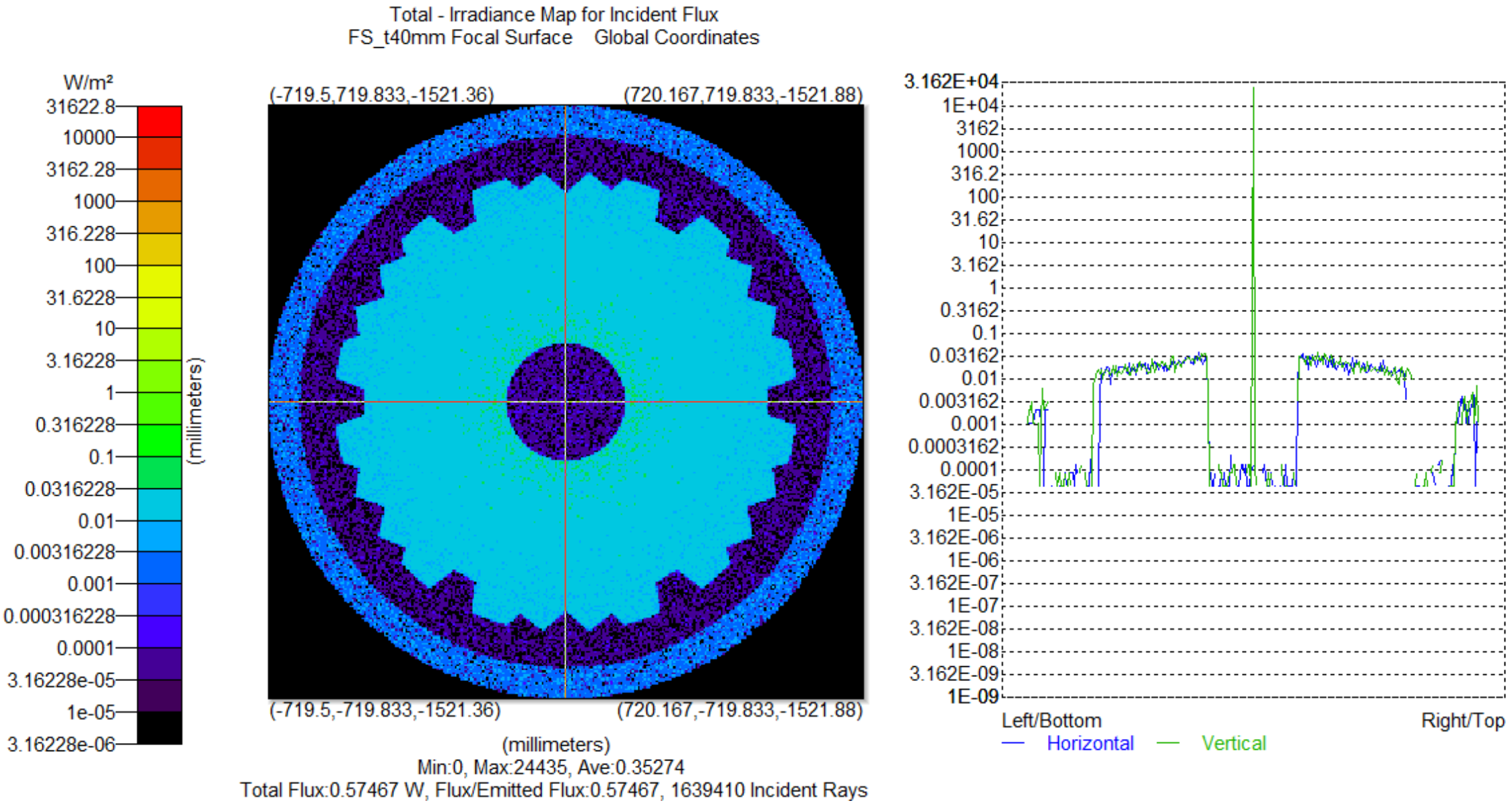

Figure 6 – Ghost image of the telescope pupil formed at the MOS focal surface for illumination with an on-axis point source. This simulation assumes un-coated field corrector lenses, flux threshold set to 0.001, and the irradiance map has 301 pixels. The cross hair is centred on the location of the in-focus image at the exact centre of the field of view.

Functional characteristics of the telescope optical components, as required for wavefront control, are briefly outlined in section 4 herein.

Possible design improvements[18] have been identified and recorded. These could most usefully be explored in phase B. Would a slightly larger secondary mirror still allow for sufficient wind rejection within acceptable technological risk and cost, the focal ratio of the primary mirror could be slightly relaxed without affecting the telescope length. This would help reducing the MOS field aberrations. The same applies if the segment fabrication metrology would allow for a more complex segment prescription (asphere instead of pure conic). A minor re-optimisation of the IFS path, implying only a small change of the position of the altitude axis, would improve margins – currently deemed low – with respect to optical beam volumes near the MOS focal surface. Short design runs also give confidence that the central obscuration of the IFS beams, currently driven by the mirror M7 and slightly larger than at the MOS focus, could be reduced and made equal to the MOS one, without loss of optical quality. Atmospheric Dispersion Compensation using different degrees of freedom within the field corrector could also be explored.

## 3. STRUCTURE AND ENCLOSURE

The alt-azimuthal telescope (Figure 7) will be enclosed in an enclosure like that of ESO's Very Large Telescope (VLT), with ventilation louvers and a permeable wind screen shielding the telescope against excessive wind buffeting. Although the telescope pier needs to be made relatively high to accommodate the nearly 200 integral field spectrographs in a gravity-stable configuration, the compact design of the telescope, with an f/0.95 primary mirror, allows for dimensions (Ø 37-m, height 38-m) "only" 30% larger than that of the VLT enclosure.

The design and analysis effort is reported elsewhere in expanded form[19],[20]; herein we only provide a summary.

The altitude moving mass is estimated at 130 tons, and the total telescope moving mass at 600, including the MOS modules – with an allocation of 200 tons. Several configurations are being explored for the location of the modules: distributed on the azimuth floor or partially embedded into it. As these modules represent considerable volumes, in all cases they are placed behind the telescope to avoid disrupting wind flow in front of it. Essential criteria for the ongoing trade-off include fibre length, access, structural stiffness, earthquake safety. Free space below the primary mirror is also reserved for integration of the fibre front-end through the hole of the primary mirror.

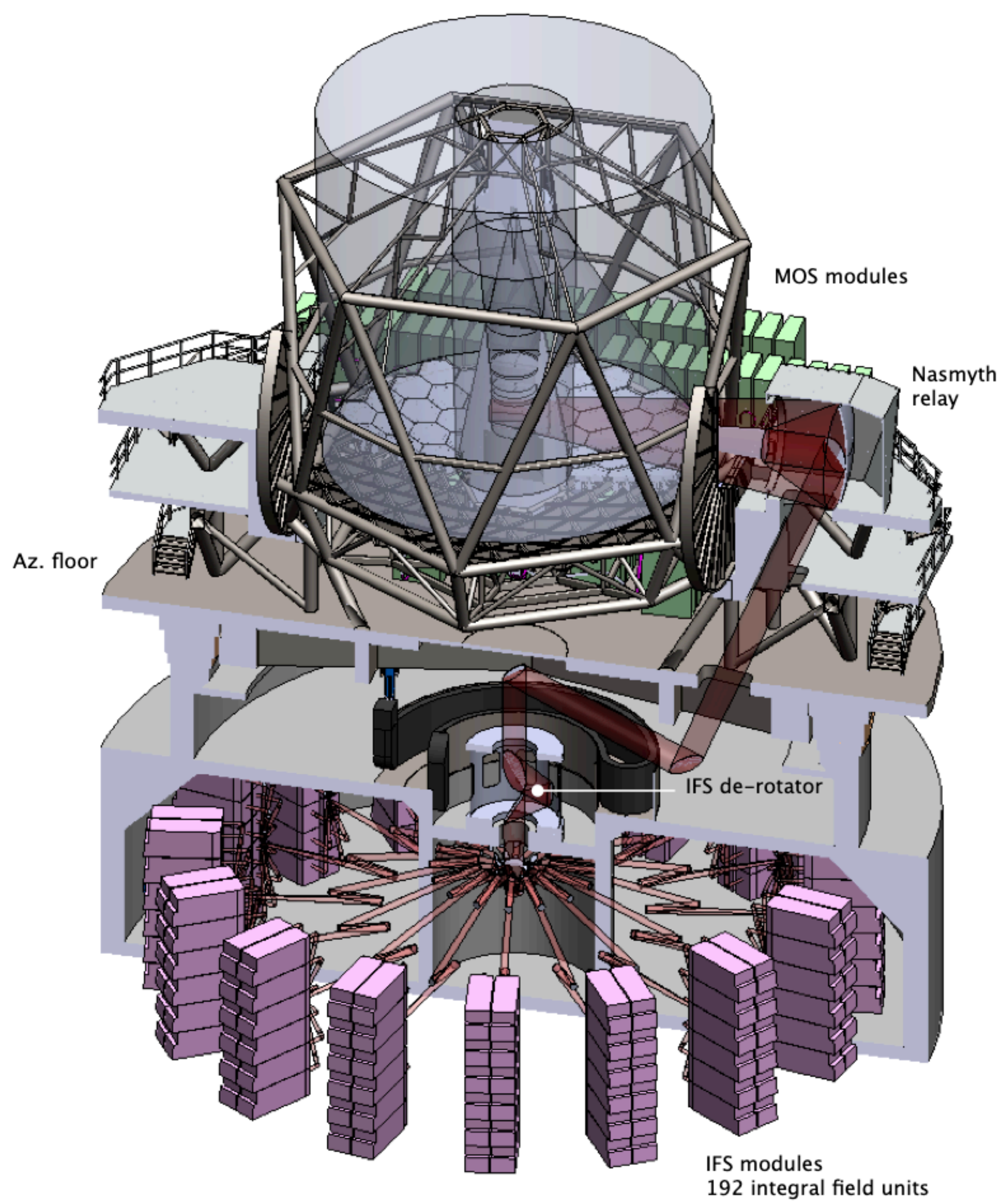

Figure 7 – Overall layout

The altitude axis is 9-m above the azimuth floor. Of the two Nasmyth platforms, one is reserved for the Nasmyth group of the IFS relay optics. This group includes the three mirrors M6, M7, and M8. An ongoing trade-off focusses on the mounting of the convex mirror M7 – the secondary mirror of the Cassegrain assembly on the Nasmyth platform. This mirror M7 is adaptive and mounted in a cell allowing rigid body motion along 5 degrees of freedom. It needs to be mounted on spiders; to avoid differential rotation with the shadow of M2 spiders, M7 spiders need either to connect to the altitude structure or to a barrel supported on the Nasmyth platform and clocking the M7 spiders synchronously with the altitude structure. The trade-off is ongoing and will have to account for the differential flexures between the altitude axis and the Nasmyth reference frame.

The second Nasmyth platform has been proposed as a possible location for some MOS modules, but this option does not shorten fibre length and is no longer considered. Instead, it could be used as convenient access when extracting and integrating segments – the altitude structure includes a foldable bridge for the handling of the segments, à la ELT. The assumption is that the ELT segment manipulator and local coherencer (LOCO) would be available for integrating WST segments and coherencing them. More generally, the WST segments population representing 10% of the ELT one, we presume that WST would represent an acceptable overhead on the ELT maintenance infrastructure. Segment carts, transport and storage containers set aside, duplication of ELT tools would be the exception rather than the rule.

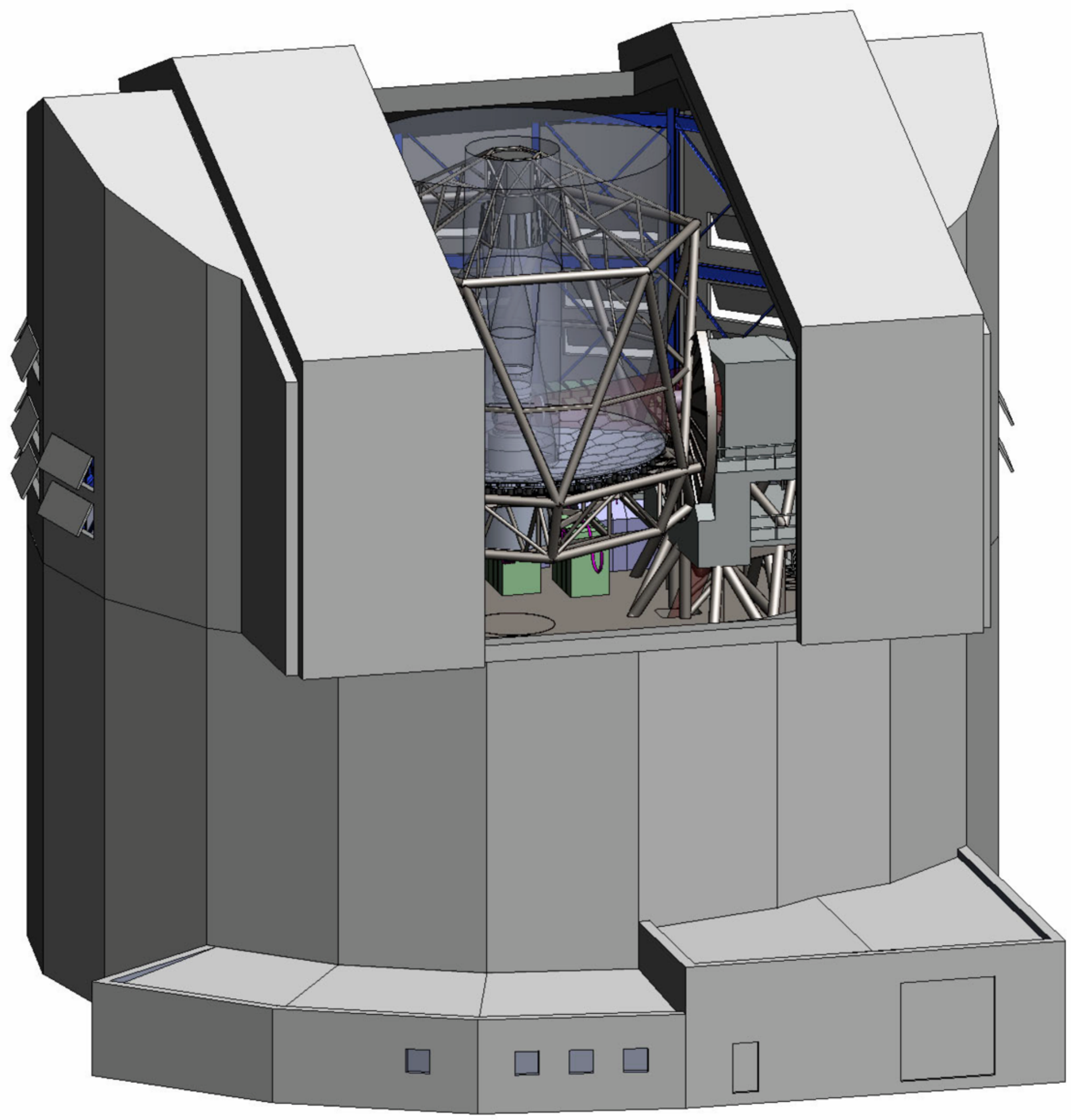

Figure 8 – WST telescope & enclosure

The first three eigenmodes (clocking of the secondary mirror and lateral motion of the structure) are shown in Figure 9; the third (locked rotor mode) is estimated at ~8.5 Hz. This figure is probably optimistic since some mass allocations still need to be refined; conversely, finite element optimisation of the structural design has just started. Highest von Mises stresses range between ~100 and 120 MPa, depending on zenithal distance. The quasi-static differential rigid body motion between the primary and the secondary mirror is in the range 0.8 mm lateral and axial, from zenith to horizon.

Computational Fluid Dynamic (CFD) modelling is applied to sanity-check the design and verify that wind flow on the telescope structure and primary mirror can be regulated. For illustration, Figure 10 shows a test case, with input wind speed 10 m/s and the wind screen is adjusted to deliver a (probably excessive) 5 m/s wind flushing of the primary mirror.

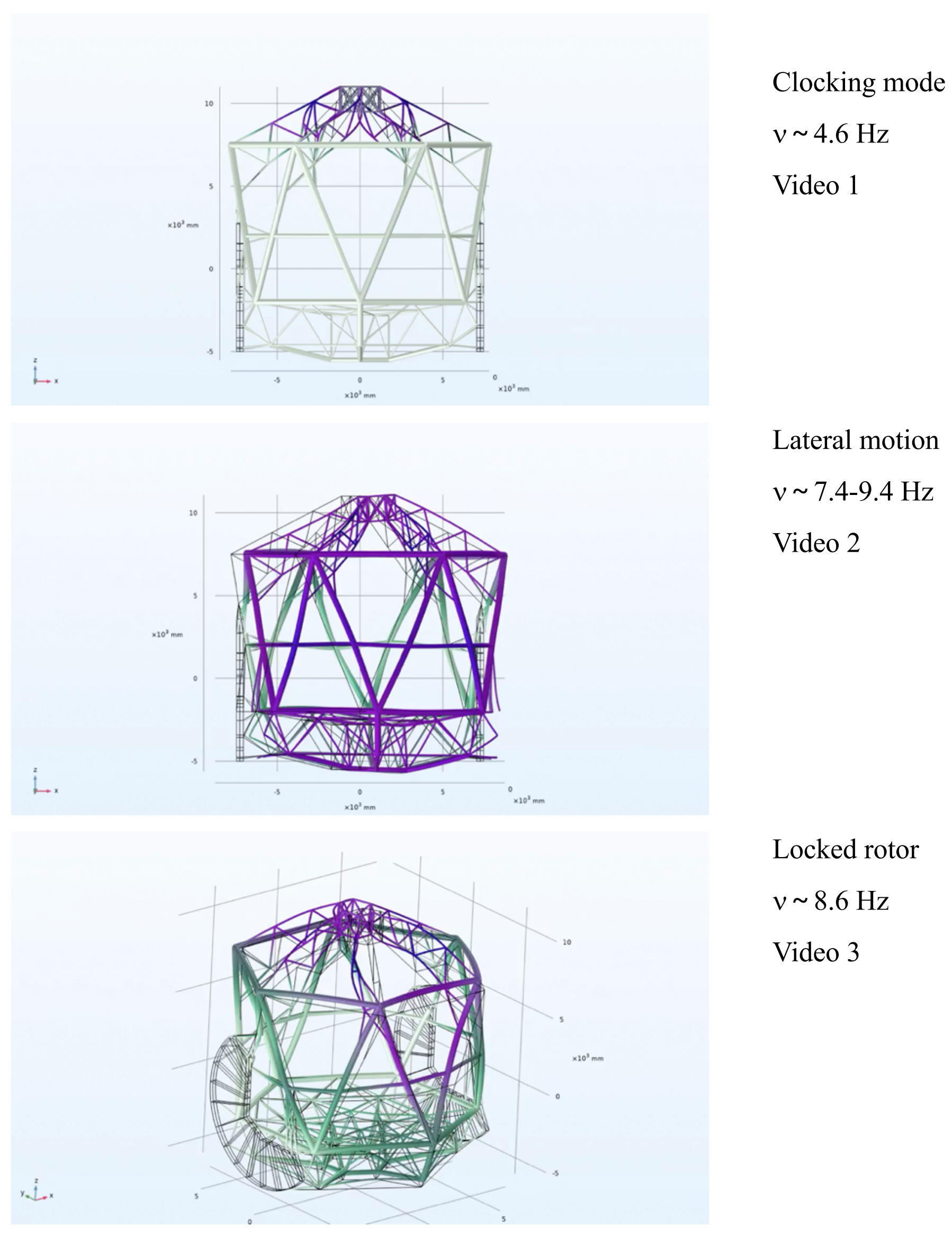
Clocking mode
ν ~ 4.6 Hz
Video 1
Lateral motion
ν ~ 7.4-9.4 Hz
Video 2
Locked rotor
ν ~ 8.6 Hz
Video 3

Figure 9 – WST structural design, first eigenmodes. **http://dx.doi.org/doi.number.goes.here**, **http://dx.doi.org/doi.number.goes.here**, **http://dx.doi.org/doi.number.goes.here**

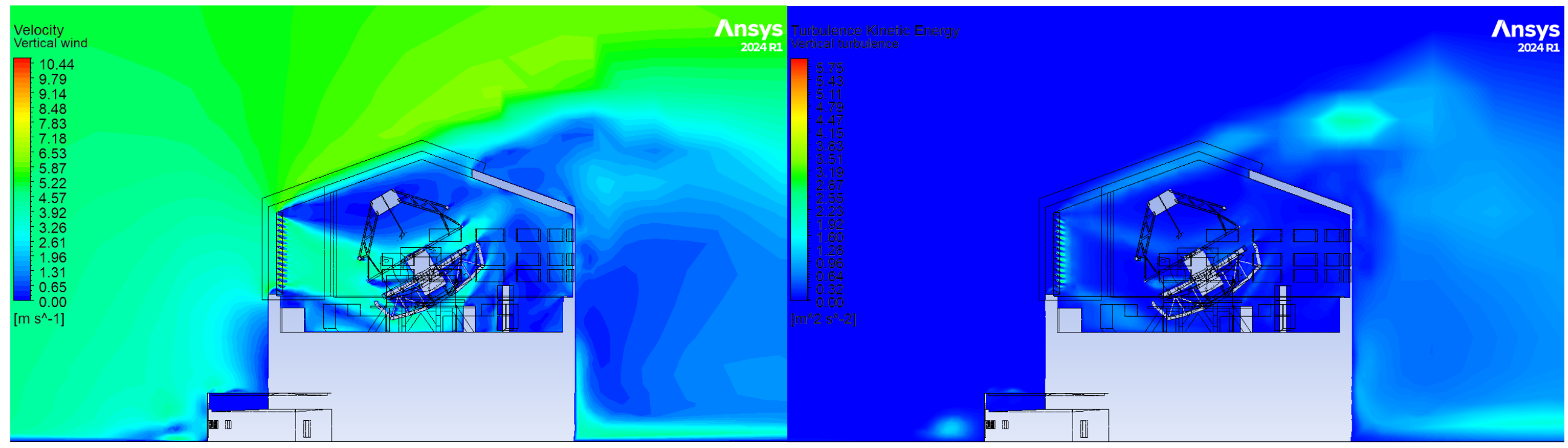

Figure 10 – CFD simulation; upfront wind speed 10 m/s, wind screen deployed and adjusted to maximise wind flushing of the primary mirror (test case). Left: wind velocity distribution; right: turbulence kinetic energy.

## 4. WAVEFRONT CONTROL

The objective of wavefront control is to acquire and update the state of the telescope at presetting, then hold it to plate scale and optical quality specification over the length of an observing block, at minimal overhead on science time. It is essential to understand that this does not necessarily imply restoring the nominal prescription – doing so could add significant and unnecessary complexity and cost, by increasing the number of required degrees of freedom and observables.

Regarding the simultaneous availability of MOS and IFS foci, the WST scenario is that the telescope is "driven" at the MOS focus, with the subsequent IFS optical path providing the degrees of freedom and metrology – à priori assumed to be sky-metrology – necessary to mitigate IFS-specific perturbations and deliver seeing reduction. An early project decision has been to set the interface between telescope and IFS instrumentation at the coudé focus, primarily for project management reasons (work breakdown structure); from a system perspective, it could have been put at the mirror M3 instead.

An initial scenario for the MOS wavefront control strategy is being elaborated – our reporting below reflects work in progress, not an established baseline. Clear functional and physical definitions of MOS sky metrology are currently evolving rapidly.

The "first order"[2] wavefront control is to ensure that line of sight and field of view, as seen by the primary mirror, project onto the distribution of fibres in a stable manner and with stable dimensions. The incentive to minimise differential motion between the reference frames of the MOS wavefront sensors and of the primary mirror is strong, since this would naturally minimise departure from the nominal paraxial properties[3] – and therefore minimise plate scale variations.

The necessary degrees of freedom include the telescope drives, the secondary mirror tip-tilt and focus stages, and motion (single-direction decentre and tilt, and piston) of a lens in the field corrector for atmospheric dispersion compensation. Atmospheric dispersion compensation (ADC) is provisionally set aside, under the assumption that the ADC is configured in open loop at presetting and left untouched for the duration of the observing block.

The upper limit of the bandwidth is set by wind shake. Observables are delivered by acquisition and guiding sensors, and by curvature sensors mounted off-axis on the fibre front-end. They are complemented by look-up tables.

The second order shall

- ensure that the primary mirror remains continuous in amplitude and first order spatial derivative, and
- mitigate quasi-static, field-independent wavefront perturbations to the possible extent.

The available degrees of freedom include:

- the primary mirror control system – segments piston, tip-tilt (position actuators), figure (warping harnesses); local metrology (capacitive or inductive edge sensors). The edge sensor – position actuator close loop bandwidth is about 2 Hz (ELT specification).
- the secondary mirror 5-degrees of freedom rigid body motion, capable of relative high bandwidth for tip-tilt (a few Hz).

Rigid body tip-tilt motion of the entire fibre front-end is also being considered, in case the secondary mirror commands would lead to unacceptable tilt of the large focal surface. With three sets of curvature sensors off-axis mounted onto the fibre front-end, this tilt would be observable and could feed into the control strategy. In view of the added complexity however, this additional set of degrees of freedom is kept as backup only, subject to further analysis.

**Primary mirror** – The edge sensors of the primary mirror need to be calibrated on-sky periodically. This is done on-sky using established techniques[13] at a frequency of once every 2-4 weeks, as set either by the drift of the sensors or by the recoating of segments – if required. The calibration technique implies perturbing the state of the mirror and can therefore not be applied concurrently with science observations. Phasing runs, therefore, should be able to accommodate poor seeing conditions, good ones being prioritized for science. WST would be equipped with a phasing and diagnostic camera deployed-on axis at the MOS focus – there is space for a phasing-shaping camera mounted on a swing-arm, either using M3 as pickoff or with a pickoff mirror above M3 viewing the on-axis focus when deployed. Coherencing, if needed, would be done during daytime, using the ELT LOcal COherencer (LOCO), prior to a phasing run. Based on experience gathered at during ESO's test runs at the Gran Telescope Canarias (GTC), the overhead on science time is estimated at about half an hour if the mirror is coherenced prior to the phasing run, about 2 hours if not.

---

[2] Our terminology, "first" and "second" order loosely refers to a spatial frequency scale and is notional only; it does not necessarily imply a hierarchical order and does not exclude interactions or inter-dependencies between "orders".
[3] The field corrector having little optical power, the paraxial properties are essentially set by the relative positions of the primary and secondary mirrors, and of the focal surface. If the focal surface and the primary mirror reference frames are aligned and rigidly tied, the paraxial solution for the position of the secondary mirror is unique.

The edge sensor-based control of the primary mirror would guarantee the continuity of the surface but is known to have poor visibility of low order modes. The edge sensors setup needs, therefore, to be complemented with low spatial order on-sky metrology. Shaping of the moderately flexible, 50 mm thick, 1.4-m segments also requires occasional adjustments.

**Secondary mirror** – WST reference design assumes a fast steering, ultra-lightweight, 2.6-m Zerodur™ secondary mirror, ~300 mm thick, with a substrate design closely matching that developed by Schott for the ELT M5[21] and a mirror cell akin to the one SENER designed for the same mirror. The WST M2 moving mass is roughly estimated at ~800 Kgs, including ~500 Kgs for the mirror substrate (first natural frequency ~200-300 Hz free-free). This is approximately twice the mass of the ELT ultra-lightweight Silicon Carbide M5 mirror. The optical beam compression on WST M2 being an order of magnitude lower than that on ELT M5, equivalent on-sky tip-tilt stroke implies an order of magnitude less stroke on WST M2 than on ELT M5. According to simulations performed by SENER, the ELT M5 cell, mated with WST M2, would thus deliver ELT-like performance despite the higher mass and inertia of the WST mirror.

An active meniscus option on actively damped pneumatic support is as an alternative, with the advantage of providing some active correction capability – with a spatial frequency limited, however, by the large beam displacement on the mirror.

**MOS sky metrology** – Several configurations of the wavefront sensors are being explored, combining lessons from existing facilities (e.g. VISTA[22]) and ongoing projects – with a combination of curvature sensors (à la 4MOST[23]) for low order modes, and Shack-Hartmann for stacking[4] and shaping of the segments.

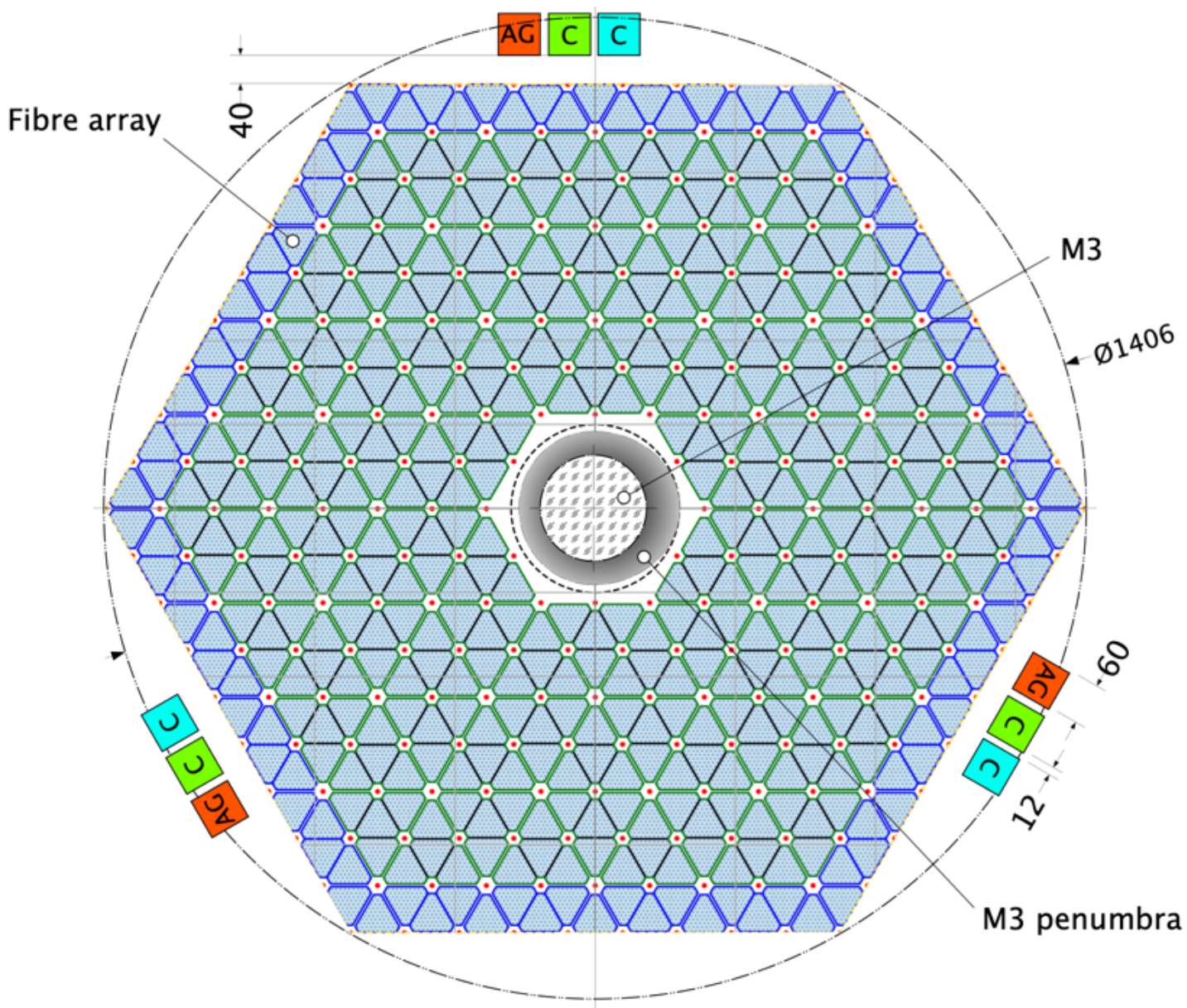

Figure 11 – Distribution of curvature and acquisition & guiding sensors. "C" refers to the low order curvature sensors, "AG" to acquisition and guiding ones. The distribution of fibre modules is shown for illustration only and does not necessarily represent the baseline accurately. Shack-Hartmann sensors are not shown.

A possible arrangement of the curvature sensors is shown in Figure 11, with three sets of curvature sensors and acquisition and guiding cameras mounted on the MOS front-end. Shack-Hartmann are not shown as implementation in the design has just started. A dummy front-end is required, with duplicate sensors – to allow telescope operations before the science front-end becomes available or is removed from the telescope. Another configuration would be to mount the MOS wavefront sensors on a structure mounted onto the Cassegrain de-rotator and cross-referenced to fiducials attached to the MOS front-end[5].

---

[4] In our terminology, "stacking" refers to stacking the image spots given by the individual segments – i.e. rigid body tip-tilt of the segments, <u>not</u> inter-segment piston steps. As such, it includes low order modes of the overall mirror.

Building on 4MOST characteristics, and considering the WST's larger collecting area, we estimate that 60 x 60 $mm^2$ sensors, providing 25 arc $minutes^2$ each, will suffice to guarantee ~99.5% sky coverage. With f/3.3 input beams and a ±4 mm defocus of the curvature sensors, we expect to capture a sufficient range of spatial frequencies to mitigate the lowest order modes of the primary mirror, most poorly visible to the edge sensors.

The curvature sensors need to be supplemented with higher order ones, e.g. Shack-Hartmann capable of sampling the figure of the segments and, most probably, several pupil masks allowing for e.g. 3, 19 sub-apertures per segment. The expectation, to be confirmed once the ELT enters operation, is that the segments themselves will be sufficiently stable to warrant occasional re-shaping only. At the lower end of the spatial frequency range, these sensors would overlap with the curvature ones or complement them to better sample wavefront slope near the edge of the pupil. Because of vignetting off-axis (Figure 12), at least two sensors are required to reconstruct the primary mirror. Our starting assumption is that 3 Shack-Hartmann ones, with ~50x80 $mm^2$ patrol range (~ 25 arc $minutes^2$) would be co-mounted on the MOS front-end, together with the curvature ones. Again, duplicates would have to be included in the dummy front-end.

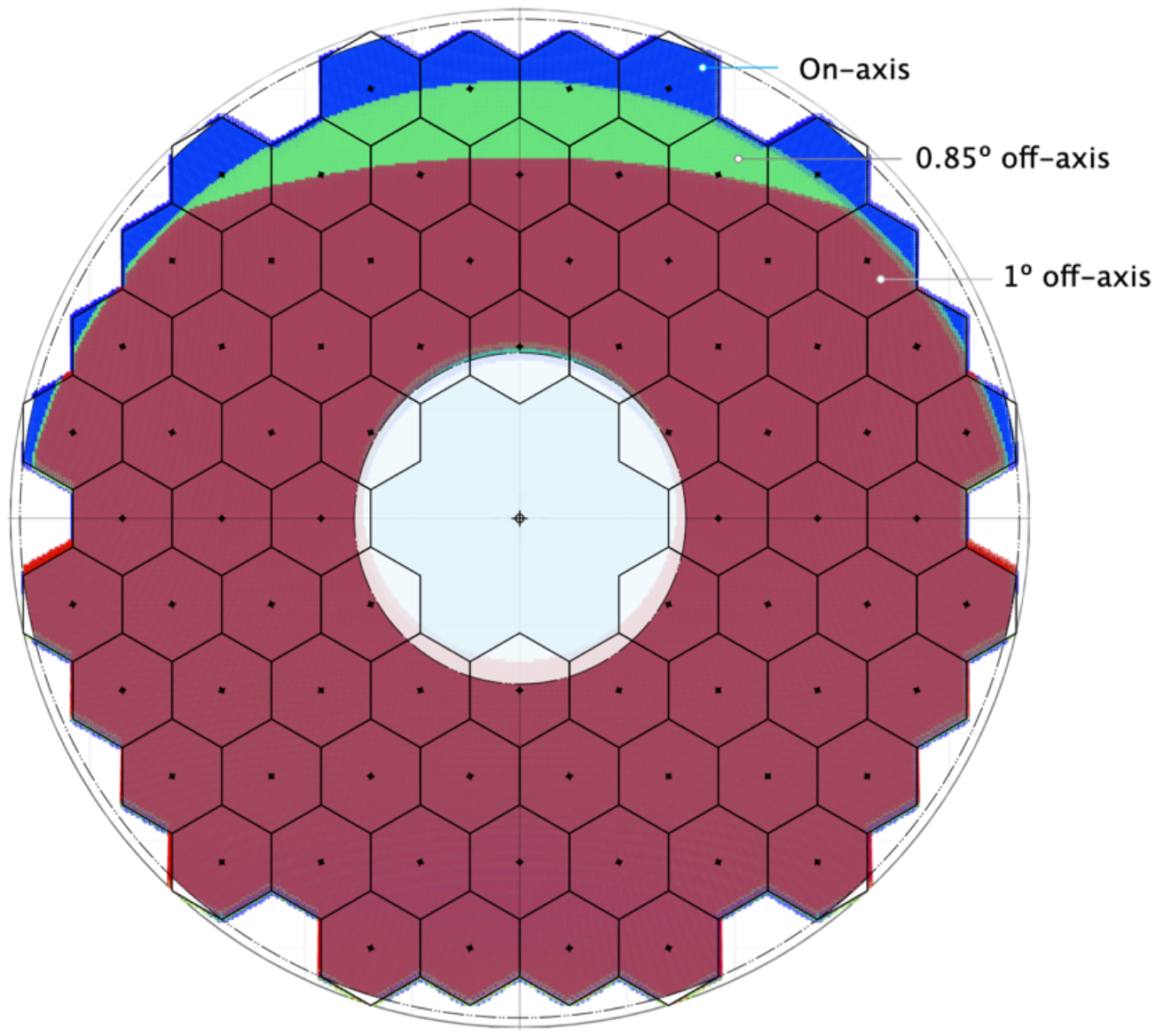

Figure 12 – Vignetting at the edge of the MOS field (cat's eye).

At the time of writing of this article, the design of the IFS-specific wavefront control scheme has not started. The 1.1-m convex M7 mirror is identified as the seeing reducer and ideally located to provide focusing and compensation of decentring aberrations between the altitude-borne optics and the Nasmyth re-imager. We will also reserve a minor fraction of the adaptive stroke to

- further improve optical quality at the IFS focal surface by compensating asymmetrical, near-constant wavefront terms across off-axis IFS subfields, and
- mitigate the effect of thermal gradients in the coudé path.

If pupil motion and/or tilt of the IFS focal surface turn out to exceed tolerances – e.g. because of runout of the optical de-rotator or of misalignment of the altitude-exiting beams and Nasmyth re-imager –, coordinated motion of selected folding mirrors in the coudé relay should à priori allow compensation.

For the sky metrology, there is ample design space available near the IFS telecentric focal surface. Simulations aimed at estimating the possible performance and at deriving key mirror and metrology specifications have started.

[5] This option, however, implies a schedule link between MOS and IFS since said cross-referencing at factory integration of the MOS front-end would be needed to enable wavefront control.

## 5. CONCLUSIONS

15 months into phase A, the conceptual design of the Wide-field Spectroscopic Telescope did not reveal any showstopper nor any worrisome increase of difficulty or complexity, when compared to existing facilities. Much work remains to be done, however, to fully appreciate the performance potential, the risks, to consolidate the design, derive robust cost estimates and establish a prospective project plan.

## ACKNOWLEDGEMENTS

This project has been funded by the European Union's Horizon Europe research and innovation programme under grant agreement No. 101183153.

The Spanish company SENER kindly provided performance estimates of its fast-steering ELT M5 mirror cell design, when mated with a moving mass twice the mass and inertia of ELT M5.